\documentclass[aps,amsfonts,prd,nofootinbib]{article}

\title{Superspreaders and  High Variance Infectious Diseases}

\usepackage[utf8]{inputenc}

\usepackage{graphicx}
\graphicspath{ {./Images/} }
\usepackage{wrapfig}
\usepackage{amsmath, amsthm, amssymb}
\usepackage{hyperref}
\hypersetup{
    colorlinks=true,
    linkcolor=blue,
    filecolor=magenta,      
    urlcolor=cyan,
}
\usepackage{theoremref}

   \newtheorem{thm}{Theorem}[section]

   \newtheorem{claim}[thm]{Claim}

\newcommand\abs[1]{{\left\lvert{#1}\right\rvert}}

\usepackage{soul}
\usepackage{xcolor}

\usepackage{mwe}
\usepackage{float}
\usepackage[font=small]{caption}
\numberwithin{equation}{section}

\usepackage{fullpage}
\usepackage{authblk}

\author[1]{Yaron Oz}
\author[2]{Ittai Rubinstein}
\author[2]{Muli Safra}

\affil[1]{Raymond and Beverly Sackler School of Physics and Astronomy, Tel-Aviv University, Tel-Aviv 69978, Israel}
\affil[1]{School of Natural Sciences, Institute for Advanced Study, Princeton NJ, USA}
\affil[2]{Blavatnik School of Computer Science, Tel-Aviv University, Tel-Aviv 69978, Israel}

\date{\today}
\usepackage{theoremref}

\begin{document}
\maketitle
\begin{abstract}
A well-known characteristic of pandemics such as COVID-19
is the high level of transmission heterogeneity in the infection spread:
not all infected individuals spread the disease at the same rate and some individuals (superspreaders) are responsible for most of the infections. To quantify this phenomenon requires the analysis 
of the effect of the variance and higher moments of the infection distribution.
Working in the framework of stochastic branching processes, we derive an approximate analytical formula for the probability of an outbreak in the high variance regime of the infection distribution, 
verify it numerically and analyze its regime of validity in various examples. We show that it is possible for an outbreak not to occur in the high variance regime even when the basic reproduction number $R_0$ is larger than one and
discuss the implications of our results for COVID-19 and other pandemics.
\end{abstract}

\section{Introduction}

The classic SIR models provide an epidemiology framework for studying 
the spread of a disease~\cite{R0}. 
The basic reproduction number $R_0$ in these models is the mean value of secondary infections caused by an infected individual. It determines the threshold $R_0>1$ for an outbreak.
Alternatively, it determines the fraction of the population that will be infected before herd immunity is reached.
In view of the importance of this parameter, major measures (such as lockdowns) are taken in order to reduce the value of $R_0$. The estimation for the COVID-19 pandemic, for example,  is  $R_0\sim 2-3$.

The structure underlying the epidemic spreading is that of a complex heterogeneous network,
where a small number of the nodes act as hubs while the majority of nodes have few contacts (for a review and references therein see \cite{review}).
Indeed, not all people cause a similar number of secondary infections and there is clear empirical evidence for high levels of transmission heterogeneity in the infection spread (see e.g. \cite{NatureCovid2,Adi,science}).
The analysis in \cite{Adi}  for the COVID-19 pandemic suggests that between $5\%$ to $10\%$ of infected individuals
are responsible for $80\%$ of secondary infections.
This may be due to differences in the number of contacts, in protective equipment, in levels of hygiene, in time of diagnosis or biological effects such as tendency to cough and sneeze. 

Individuals with high secondary infection rate are commonly referred to as {\em superspreaders}.
This is encoded in the degree distribution of the epidemic spread network corresponding to the 
{\em  infection distribution}. While homogeneous random networks such as the Erdos-Renyi model exhibit a 
statistical homogeneity of the nodes and the degree distribution is peaked around the average value,
heterogeneous networks such as the scale free models reveal a power law structure of the degree distribution and nodes with very large degree. 

The infection distribution is taken not over a random individual, but rather over a random infected individual, i.e. it is weighted according to the \textit{a priori} probability of each individual to be infected.
For instance, an individual in contact with many people has a higher likelihood both to be infected and to infect others and this is reflected in the degree of the corresponding node in the network.

Studying the phenomenon of superspreaders, which seems  to follow the Pareto-type Principle \cite{2080}, as well as its implications on the spread of the disease is crucial when devising and implementing control policies \cite{2080,hetero}.
In order to analyze the impact of the superspreaders on the epidemic spread we have
to consider the effect of the variance and the higher moments
of the infection distribution.
The main goal of this paper, is to study a question of utmost importance when facing pandemics such as COVID-19, namely:``what is the probability that a disease will disappear without a major outbreak?”

An outbreak is often referred to as a sudden rise in the number of infected individuals. In this paper, however, we define an outbreak with reference to the total fraction of infected individuals in the long term and not at any specific point in time.
Thus, we consider that {\em an outbreak has not occurred} if the disease has disappeared with a negligible herd immunity. 
Note, that we will analyse the natural evolution of the disease irrespective of the measures---social and others---taken to reduce $R_0$.

We will work in the framework of  Galton-Watson branching processes (for a review see e.g. \cite{GW}), and use it to predict the probability of an outbreak as a function of the infection distribution, that is the probability distribution for an individual to infect a given number of people.
We derive an approximate analytical formula for the probability of an outbreak in the high variance regime of the infection distribution,
verify it numerically in various examples, compare it to COVID-2 data and discuss its implications for the COVID-19 pandemic.
In particular, we will show that it is possible for an outbreak not to occur in the high variance regime even when the basic reproduction number $R_0$ is larger than one. This phenomenon has been observed in numerical simulations \cite{Barzel}.

\section{The High Variance Regime}

The infection distribution specifies, for each natural number $k$, the probability of an infected individual to infect $k$ others.
We denote by $R_0$ and $V$ the mean and variance of the number of people infected.
When $R_0 < 1$, it is well established that the disease would disappear on its own, while
when $R_0 - 1$ is not small compared to the variance $V$, one can use deterministic models such as SIR that provide an accurate description.

Let us thus focus on the high variance regime:
\begin{equation}
0 < R_0 - 1 \ll V \ .
\label{regime}
\end{equation}
Our main result can be stated as follows.
The probability that a disease will disappear without herd immunity is:
\begin{equation}
{\sf Pr}=\gamma^n, 
\label{Analytical1}
\end{equation}   
where $n$ is the current number of infected individuals 
    and $\gamma$ in the regime (\ref{regime}) can be approximated as:
    \begin{equation}
\gamma
     \approx 1 - {\cal Q} \ ,
     \label{Analytical2}
\end{equation}
where
\begin{equation}
{\cal Q} = \frac{2(R_0 - 1)}{R_0 ^ 2 + V - R_0} \ . \label{Q}
\end{equation}
Below, the corrections to the approximate formula (\ref{Analytical2}) and (\ref{Q}) are
bounded by higher powers of the ratio ${\cal Q}$ as well as the higher moments of the infection distribution.

In section~\ref{section:formal} we  formulate the main result precisely and prove it.
However, before delving into the proof let us consider some of its qualitative implications,
compare it to pandemic data and numerically verify its accuracy.
First, the larger the variance $V$ compared to $R_0-1$, the higher the probability for the disease to disappear  before herd immunity is reached. Thus, the fate of the disease does not depend only on $R_0-1$.
Second, the fewer infected individuals, the
higher the probability for the disease to disappear and, consequently,
the less stringent the pandemic measures that must be taken, even when $R_0 > 1$. 
Third, the effective dependence on the variance is $\frac{V}{n}$.

Let us numerically compare our approximate analytical formula to the exact $\gamma$
for the re-scaled infection distribution of COVID-2~\cite{NatureCovid2}.
The latter is based on fitting pandemic data to a distribution obtained by sampling a Poisson distribution whose mean is sampled from a Gamma distribution, which we will call Gamma-Poisson distribution.
Since $R_0$ of COVID-2 is high, we define the infection distribution for lower values of $R_0$ by re-scaling the original one, that is, we fix the shape of the distribution that is determined by a parameter $k$ and re-scale
the parameter $\theta$ that determines the scale of the distribution.
The results are depicted in figure 1 and, as expected, we see that lower $R_0$ implies better accuracy.

\begin{figure}[H]
\centering
\label{fig:gamma_estimate}
\includegraphics[width=.8\columnwidth]{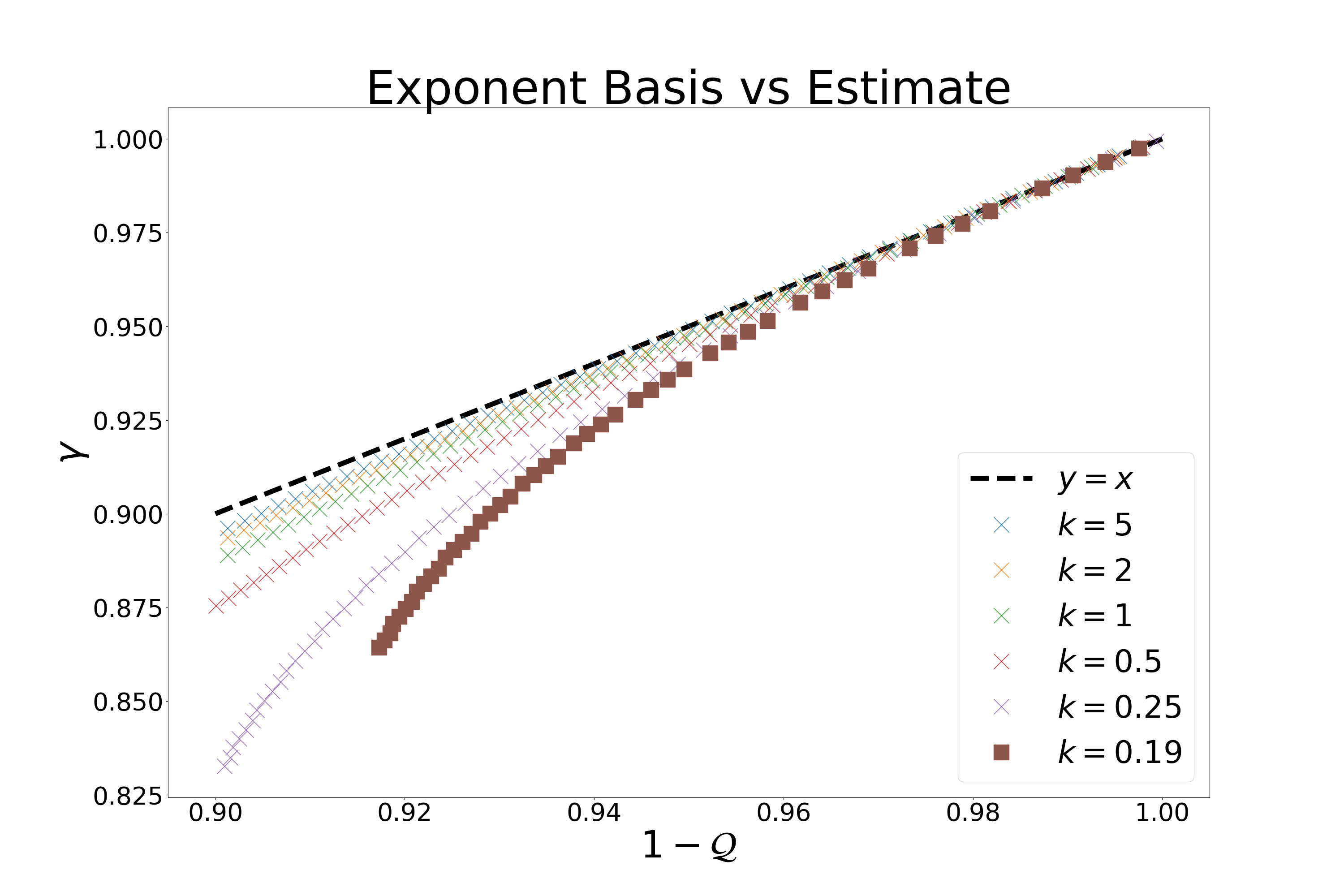} 
\caption{
    Our approximate formula $1- {\cal Q}$ vs. the exact Galton-Watson coefficient $\gamma$ for the re-scaled infection distribution of COVID-2~\cite{NatureCovid2}. The latter fits data to a distribution obtained by sampling a Poisson distribution with mean being sampled from a Gamma distribution. The parameter $k \approx 0.19$ for COVID-2 controls the shape
    of the Gamma distribution, while the parameter $\theta$  controls its scale. We constructed our data by fixing $k$ and scale $\theta$ to give different values of $R_0=k\theta$. We reach $R_0\approx 1.6$ for the lower value of $k$.
    }
    \vskip-1em
\end{figure}

We use our formula to estimate the probability to avoid an outbreak for the 
COVID-2 and COVID-19 pandemics as a function of $R_0$ and the number of infected $n$---requiring an estimate of the ratio $\frac{V}{R_0^2}$.
Based on \cite{NatureCovid2} we set $V=5R_0^2$ for COVID-2.
As noted above, the analysis in \cite{Adi} estimates that the $p_h$ value 
(the percentage of the infected population responsible for $80\%$ of all secondary infections) 
is around $5\%-10\%$. Assuming Gamma-Poisson distribution, $p_h=10\%$ implies $k=0.1$ and $V =10R_0^2$, and lower $p_h$ values correspond to even higher variance.
We plot the results in figures 2 and  3: for given values of $R_0$ and $n$, the higher the variance the higher the probability of avoiding an outbreak.

\begin{figure}[H]
\centering
\label{fig:out_break_prob}
    \vskip-1.8em
\includegraphics[width=0.7\columnwidth]{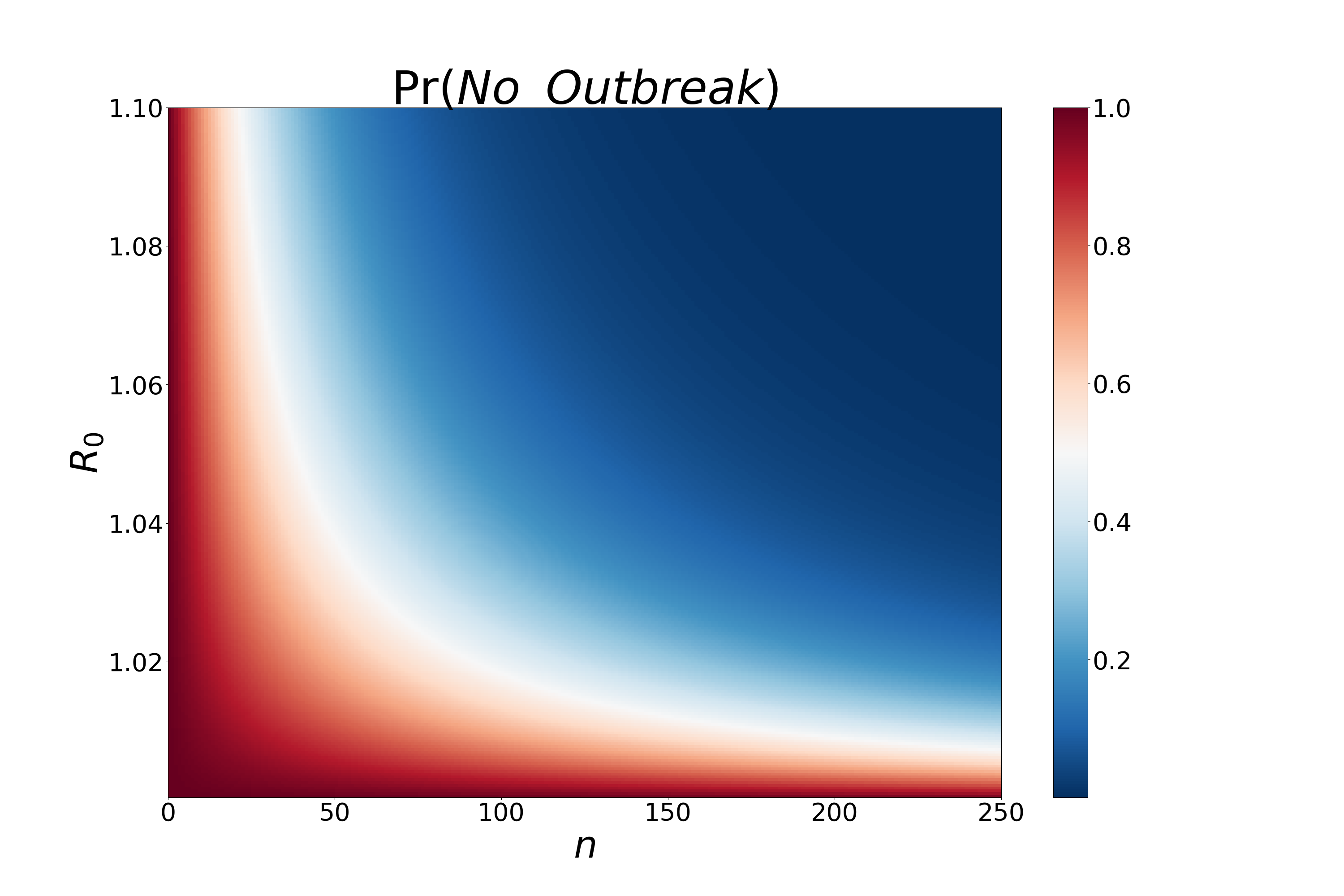} 
\caption{The probability to avoid an outbreak when $V=5R_0^2$ (COVID-2) as a function of the basic reproduction number and the number of infected individuals.} 
\includegraphics[width=0.7\columnwidth]{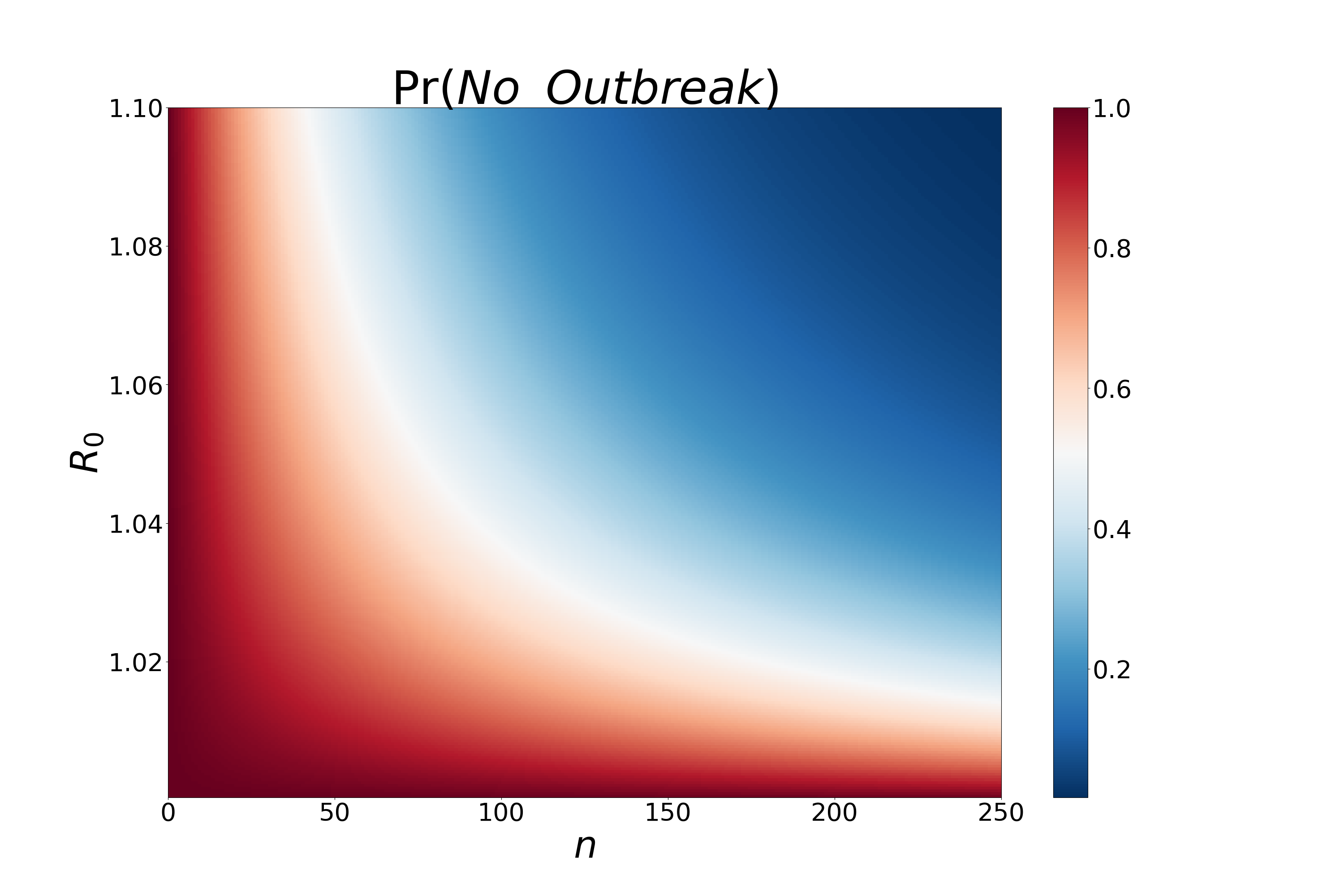}
\caption{The probability to avoid an outbreak when $V=10R_0^2$ (COVID-19) as a function of the basic reproduction number and the number of infected individuals.} 
    \vskip-1.8em
\end{figure}

While it is clear that $\gamma$ cannot be determined precisely by $R_0$ and $V$ alone and the information about the higher moments of the infection distribution is necessary, our numerical analysis reveals that 
for certain distributions that are often being employed for real world pandemics the accuracy of (\ref{Analytical2}) 
is mostly determined by the value of $R_0 - 1$ as depicted in figure 4 and figure 5.

In figure 4, the Poisson distributions  has $\lambda$ values in the range $1.0$ to $1.1$,
the geometric distributions has $p$ values in the range $0.43$ to $0.5$,
the Poisson$10x$ distribution is obtained by selecting a Poisson distribution with $10 \leq \lambda \leq 20$ value with probability  $10\%$ or the zero distribution with probability $90\%$, and the Truncated Power Law distributions has a cut-off at $100$ with powers ranging from $2.1$ to $2.375$.
In figure 5 we consider the ratio between the logarithms since this determines the ratio between the values of $n$ that would give a specific probability to avoid an outbreak.

\begin{figure}[H]
\label{fig:other_dists}
\centering
\includegraphics[width=0.7\columnwidth]{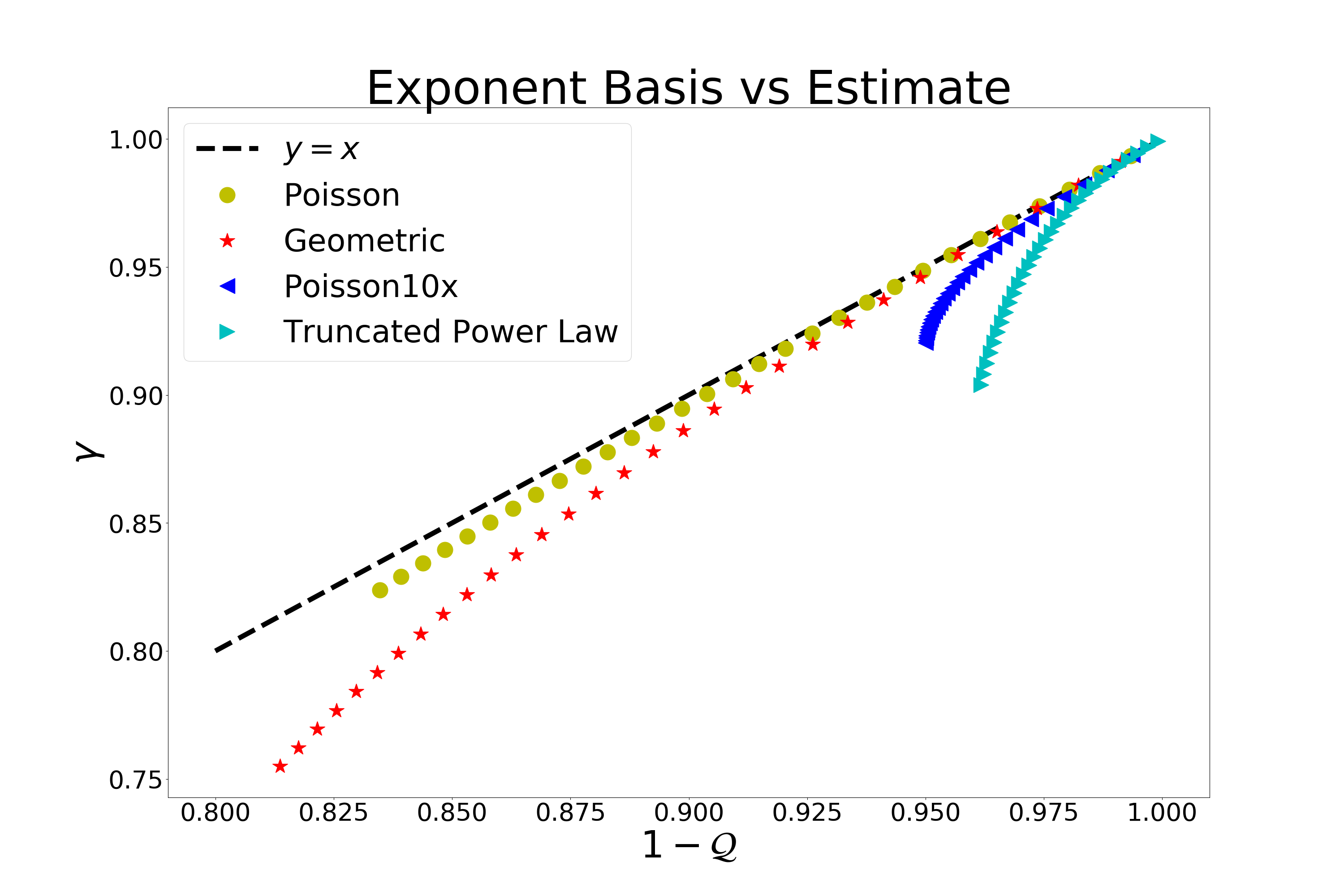} 
\caption{The exact Galton-Watson coefficient $\gamma$ vs.
    our approximate formula $1-{\cal Q}$ for various infection distributions. The line $y=x$ corresponds to $\gamma=1-{\cal Q}$.
    The accuracy of the formula depends mostly upon the value of $R_0$, which explains
    the different deviations of the distributions from the $y=x$ line:
    $R_0$  for the Poisson, Geometric, Poisson10x and the truncated power law distributions are in the ranges $[1, 1.1],\; [1, 1.32],\; [1, 2],\; [1, 1.72]$, respectively.}
    

\end{figure}



\begin{figure}[H]
\label{fig:estimate_R0}
\centering
\includegraphics[width=0.7\columnwidth]{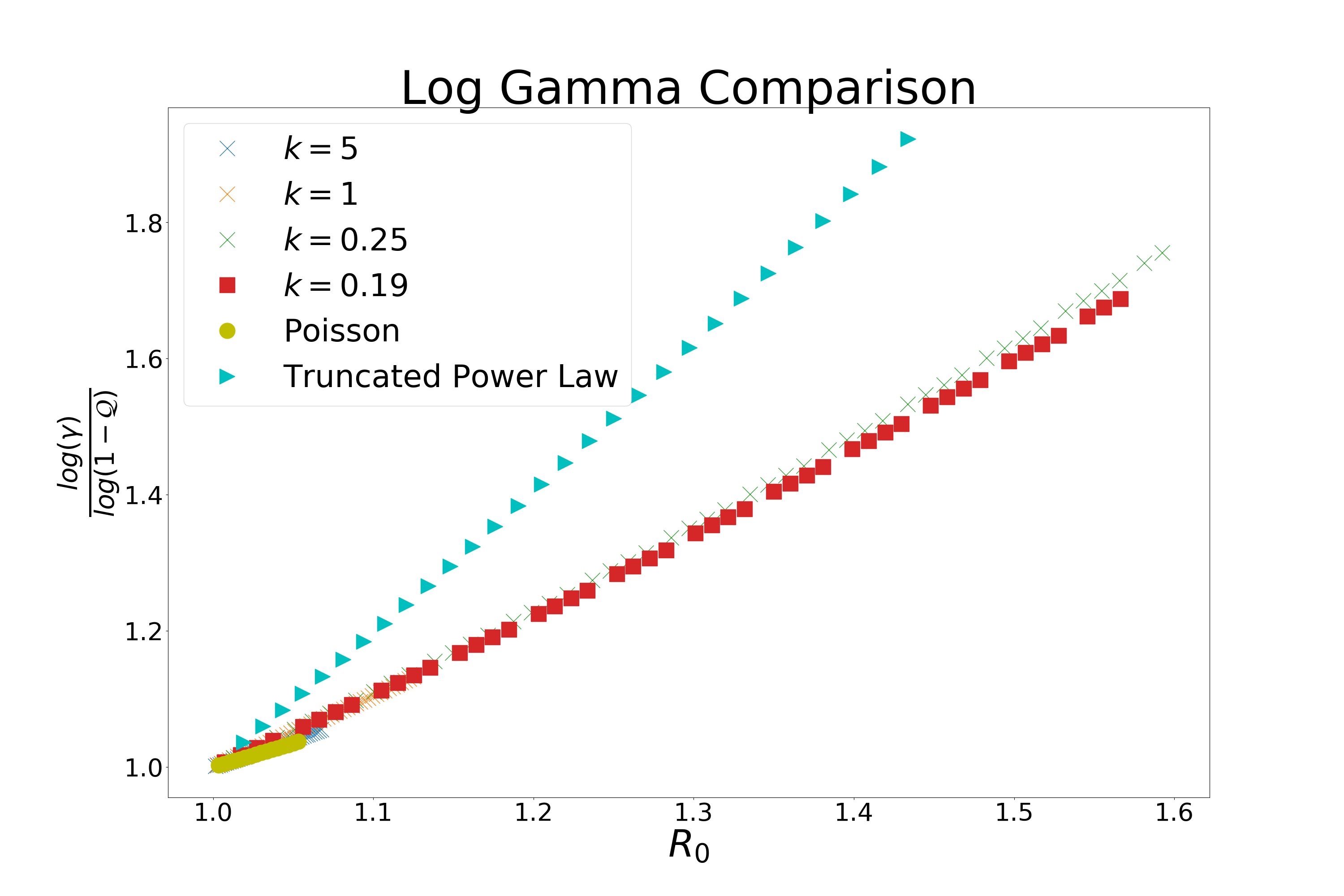}
\caption{
    A comparison of our formula to the exact value of $\gamma$ for various distributions as a function of $R_0$, and the larger $R_0$ the larger the deviation.
    We tested our results for several distributions: Gamma-Poisson distributions for COVID-2 \cite{NatureCovid2},  Poisson and Truncated Power Law distributions. 
}
\end{figure}

\section{Formal Statements and Proofs}\label{section:formal}

We define the infection distribution by a sequence of real variables $a_k$, where $a_k$ is the probability that a carrier infects $k$ individuals and is removed. 
The normalization condition is:
\begin{equation}\label{sum}
    \sum_k a_k = 1.
\end{equation}

Denote by $M_i$ the $i$th moment of the infection distribution:
    \begin{equation}
 M_i = \sum_k {a_k k^i},~~~~M_1 = R_0,~~~~M_2 = V + R_0 ^ 2      \ ,
 \label{M}
        \end{equation}
         and the quantity $\eta$ by:
  \begin{equation}
\eta = \frac{1}{R_0 ^ 2 + V - R_0} \sum_{i \geq 3} {\left(\frac{3{\cal Q}}{2}\right) ^{i-2} \frac{M_i}{i!}}     \ .
\label{eta}
  \end{equation}
Let $p(x)$ be the polynomial 
    \begin{equation}
    p(x) =  \sum_k {a_k x^k} - x \ .
    \label{pol}
    \end{equation}

The proof of our result (\ref{Analytical1}) and (\ref{Analytical2}) consists of proving three statements:
\begin{itemize}
    \item ${\sf Pr}=\gamma^n$ where $0 \leq \gamma < 1$ is a root of $p(x)$.
    \item $p(x)$ has a root within a small neighborhood of $1 - {\cal Q}$.
    \item $p(x)$ has at most a single root in $[0, 1)$.
\end{itemize}

Let us prove the following claims:
\begin{claim} [Galton-Watson analysis] \thlabel{Galton-Watson analysis}
    If $R_0 > 1$ then the probability that the disease will disappear without herd immunity is $\gamma^n$ where $\gamma$ satisfies $p(\gamma) = 0$ and $0 \leq \gamma < 1$.
\end{claim}

\begin{claim} [Approximate Formula] \thlabel{Approximate Formula}
    There exist $c$ and $\eta_0 > 0$ s.t. if $\eta < \eta_0$ then $p(x)$ has a root within the interval
    \[\left [1 - (1 - c\eta) {\cal Q}, 1 - (1 + c\eta) {\cal Q}\right]\] 
\end{claim}

\begin{claim} [Single Root] \thlabel{Single Root}
    $p(x)$ has exactly one root in the interval $[0,1)$.
\end{claim}

Combining the above assertions, we see that if:

    \begin{itemize}
        \item $R_0 > 1$ (condition for \thref{Galton-Watson analysis})
        \item $\eta < \eta_0 $  (condition for \thref{Approximate Formula})
        \item ${\cal Q} < \frac 1 {(1+c\eta_0)}$ (the root in \thref{Approximate Formula} is positive)
    \end{itemize}
then we arrive at our main result:
 \begin{equation}     
  1 - \gamma \in [1-c\eta,1+ c\eta] {\cal Q} \ .
    \end{equation}

\thref{Galton-Watson analysis} is a standard analysis of Galton-Watson processes \cite{GW}, which we will now briefly review for completeness.

One views the number of sick individuals as a Markov process, where at each point 
we pick a sick individual, add the number of people infected by him
and remove him. 
As above, $a_k$ is the transition probability from a state with $n$ sick people to a state with $k$ added infected people and one removed:
\begin{equation}
 n \rightarrow n + k  - 1 \ .
\end{equation}

Denote by $f(n)$ the probability that no major outbreak will occur at any time $t>0$ if we have at $t=0$ $n$ infected people, and define $\gamma = f(1) \in [0,1]$.

For the disease to die out, every branch that begins from one of the $n$ infected individuals at $t=0$ should disappear. Since we neglect the interaction between the infected individuals, these are independent random variables and $f(n)={f(1)}^n = \gamma^n$.

Using time independence and the total probability, one gets the recursion relation:
\begin{equation}
  f(n) =   \sum_k a_k f(n+k-1) \ .
  \label{rec}
    \end{equation}
Setting $n=1$ in (\ref{rec}) we have:
\begin{equation}
\sum_k a_k \gamma^k - \gamma = 0 \ ,
\end{equation}
that is, $\gamma$ is a root of the polynomial p(x) (\ref{pol}).

Finally, in order to complete the proof of \thref{Galton-Watson analysis}, we have to show that $\gamma \neq 1$.
This is not surprising, as we are dealing with the $R_0 > 1$ regime and setting $\gamma=1$ would make the probability of an outbreak $1 - 1^n = 0$ regardless of the number of infected at $t=0$.
In order to prove the claim, we have to show that the probability of an outbreak converges to $1$ as $n \rightarrow \infty$, but this is easy to see (for instance, by applying Chebyshev's inequality on the probability that $n$ sick will infect less than $\frac{R_0 + 1}{2} n$ individuals).

Consider next  \thref{Approximate Formula}.
It is convenient to denote $\gamma=1+\delta$ and analyze the roots of $p(x)$ :
\begin{equation}
p(1+\delta) =  \sum_k {a_k (1+\delta)^k} - (1 + \delta)  = 0 \ .
\label{p}
    \end{equation}
Expanding (\ref{p}) and using (\ref{sum}) and (\ref{M}) we get:
\begin{equation}
    p(1+\delta) = (R_0 - 1) \delta + \frac{1}{2} \left(R_0 ^ 2 + V - R_0\right)\delta^2 +~~corrections
    \ .
    \label{root}
\end{equation}
From (\ref{root}) we get the approximate  formula (\ref{Analytical2}) where
the corrections are bounded by:
\begin{equation}
 corrections~~ \leq \sum_{i\geq3} {\frac{M_{i}}{i!} \delta^{i}}  \ .
\end{equation}

Let $\eta_0 = \frac{1}{10}$ and $c = 5$.
We are interested in the case where
\begin{equation}
    \begin{aligned}
        \delta &\in\left[-1-c\eta,-1+c\eta \right]{\cal Q} \\
         &\subseteq -\left[\frac{3{\cal Q}}{2} ,\frac{{\cal Q}}{2}\right] 
        \ . 
    \end{aligned}
\end{equation}

Therefore:
\begin{equation}
    \begin{aligned}
        &p(1+\delta)=\\
         &\;\;= (R_0 - 1) \delta + \frac{1}{2} \left(R_0 ^ 2 + V - R_0\right) {\delta}^{2} \pm\\
         &\;\;\;\;\;\;\; (R_0 ^ 2 + V - R_0)\delta^2 \eta  \ ,
    \end{aligned}
\end{equation}
where we denote $X = Y \pm Z$ iff $\abs{X-Y} < Z$.
It is straightforward to see that when 
$\delta = -(1+5\eta) {\cal Q}$ we have
$p(1+\delta) \geq 0$, while when 
$\delta = -(1-5\eta) {\cal Q}$, we have
$p(1+\delta) \leq 0$.
Combining these results with the Intermediate Value Theorem, we conclude the proof of \thref{Approximate Formula}.\\

In order to prove \thref{Single Root}, consider the second derivative of $p(x)$:
\begin{equation}
 p''(x) = \frac{d^2 p(x)}{dx^2}=\sum_{k\ge 2} {k(k-1) x^{k-2}} \ ,
\end{equation}
and $p''(x) > 0$ for $x>0$. 
Thus, $p(x)$ is convex in  $\mathbb{R}^+$, and must have at most two non-negative roots.
Using (\ref{sum}) we see that $x=1$ is one of these non-negative roots.
Furthermore, $x=1$ is not a local minimum of $p(x)$, since 
\begin{equation}
p'(1)=\sum_k {k a_k} - 1 = R_0 - 1 > 0 \ ,
\end{equation}
and in particular it cannot be the global minimum for $p(x)$ in $x\in \mathbb{R}^+$.
This implies that $p(x)$ must have a negative value.

$\forall x > 1 \colon p''(x) > 0$ implies that $p'(x) > p'(1) > 0$ and hence $p(x) > 0$. 
Therefore, $p$ reaches its minimum in the $\mathbb{R}^+$ region at some point $b$, $0 \leq b < 1$. 
From the Intermediate Value Theorem, there is a point $c$,  $0 \leq c < b < 1$ such that $p(c) = 0$, and it is clearly unique, concluding our proof.

\section{Discussion and Outlook}

We
have  carried out an analysis of the stochastic spread of a disease in the high variance regime of the infection distribution. 
This allowed us to study an important characteristic of the COVID-19 and other pandemics where
 not all infected individuals spread the disease at the same rate and superspreaders are responsible for most of the infections. 
We derived an approximate analytical formula (\ref{Analytical1} and \ref{Analytical2})
for the probability to avoid an outbreak in the high variance regime (\ref{regime}) and estimated its accuracy numerically and analytically. We found out that $R_0-1$ is the main control parameter for the higher moment corrections.
Curiously, for all the distributions that we analyzed we found that $\gamma \leq 1 - {\cal Q}$, giving us an upper bound on the approximation.
We compared the formula to infection distribution data and discussed its implications for the COVID-2 and COVID-19 pandemics.

Our analysis reveals the general coarse-grained structure of the infectious diseases
irrespective of the detailed graph or network structure of the disease spread.
We studied the natural evolution of the disease 
under the assumption that 
the infection and recovery are time-independent random variables.
There are several reasons to consider the time dependence of $R_0$, $V$ and the higher moments, an obvious one being the measures, social and other, taken to reduce them.
A less obvious one is related to the time-dependent details of the disease's evolution structure. There can be a major change due to a reduction in the number of superspreaders that are removed, which leads to interesting insights about the disease spread, such as reaching herd immunity faster than previously assumed \cite{time}.

\section*{Acknowledgements}

We would like to thank Nir Kalkstein for valuable discussions on the importance of the high variance to the spread of the disease, as well as Baruch Barzel for comments on the manuscript.
The work is supported in part by the Israeli Science Foundation center
of excellence.
The work is supported in part by the European Research Council (ERC) under the European Union’s Horizon 2020 research and innovation programme (Grant agreement No. 835152), 
 ISF 2013/17, BSF 2016414  and the IBM Einstein Fellowship at the Institute for Advanced Study in Princeton.

\end{document}